\documentclass[aps,prb,twocolumn,showpacs,amsmath,amssymb]{revtex4}
\usepackage{dcolumn}
\usepackage{bm}
\usepackage{graphicx}
\usepackage{bm}
\begin{document}
\title{History dependent nucleation and growth of the martensitic phase in a magnetic shape memory alloy Ni$_{45}$Co$_{5}$Mn$_{38}$Sn$_{12}$}
\author{A. Banerjee, P. Chaddah, S. Dash, Kranti Kumar and Archana Lakhani}
\affiliation{UGC-DAE Consortium for Scientific Research, University Campus, Khandwa Road, Indore-452001, Madhya Pradesh, India.}
\author{X. Chen and R. V. Ramanujan}
\affiliation{School of Materials Science and Engineering, Nanyang Technological University, N4.1-01-18, 50 Nanyang Avenue, Singapore 639798}

\begin{abstract}
We study through the time evolution of magnetization the low temperature (T) dynamics of the metastable coexisting phases created by traversing different paths in magnetic field (H) and T space in a shape memory alloy system, Ni$_{45}$Co$_{5}$Mn$_{38}$Sn$_{12}$. It is shown that these coexisting phases consisting of a fraction of kinetically arrested austenite phase and remaining fraction of low-T equilibrium martensitic phase undergo a slow relaxation to low magnetization (martensitic) state but with very different thermomagnetic history-dependent rates at the same T and H. We discovered that, when the nucleation of martensitic phase is initiated at much lower T through the de-arrest of the glass like arrested state contrasted with the respective first order transformation (through supercooling at much higher T), then the long time relaxation rate scales with the non-equilibrium phase fraction but has a very weak dependence on T. This is explained on the basis of the H-T path dependent size of the critical radii of the nuclei and the subsequent growth of the equilibrium phase through the motion of the interface.
\end{abstract}
\maketitle
\section {Introduction}
The intriguing process of nucleation and growth of competing magnetic order in the course of first order magnetic transition gives rise to interesting physical phenomena in variety of materials of current interest \cite{dagoto,soibel,roy}.  There have been many reports in materials ranging across intermetallics\cite{chatto,sen,roy1,pallavi, kranti}, CMR manganites\cite{kranti,alok,wu,alok1,rawat,macia} and mutiferroics\cite{choi}, of the arrest of first-order transformation kinetics before its completion. Further transformation to the low temperature equilibrium phase does not take place down to lowest temperature (T) indicating that the process of nucleation and growth is inhibited. Thus, at low temperatures, a fraction of transformed low-temperature equilibrium phase coexists with the remaining fraction of kinetically arrested (KA) high temperature phase that show non-ergodic behavior and glasslike dynamic response. Effect of KA has been recently reported extensively in magnetic shape memory (MSM) materials across their first-order martensitic transition (see for example Ref.\cite{vks, ito, umetsu, jip, alok2}). The process of nucleation and growth across a martensitic transition has been studied over many decades, and these MSM materials (besides their relevance to applications) provide potentially important systems for such studies with magnetic field (H) as an additional parameter. Recently, in a significant development it is shown that the liquid to crystal transformation kinetics in monatomic Ge can be inhibited by ``magic ingredient'' of pressure \cite{bhat} and for magnetic system it is magnetic field (H) which can be used as the ``magic ingredient''. It is indeed found that H as a second control variable has a decisive role both for the magnetic first order transition and on the kinetic arrest of such transformation process\cite{chup}.  

The first order transition as well as the associated supercooling and superheating can be depicted by lines in the H-T space for the magnetic materials. Similarly, arrest of kinetics also can be represented by a line for a specific cooling rate, which can allow construction of H-T phase diagram for first order transformation combined with the process of its kinetic arrest. In real multi-element materials the accompanying quenched disorder broadens the sharp first-order transition\cite{imry}. Consequently, the transition line and supercooling as well as superheating lines broadens into bands in H-T space, consisting of a quasi-continuum of lines. Each line corresponds to a region of the sample with length scale of the order of correlation length. Following the same argument, the H-T dependent kinetic arrest is depicted as band in the phase diagram and is justified from the phenomenological studies\cite{chaddah1,wu, kranti}. The broadening helps in using the second parameter H to produce quasi-continuum of states of coexisting phases having different fractions of equilibrium phase and kinetically arrested high temperature phase at low temperature, which vary with time, indicating metastability\cite{chatto,alok,alok1,wu, chaddah2}.

Thus, according to the phenomenological phase diagram, it becomes possible by traversing different H-T paths to initiate nucleation of the martensitic phase at different temperatures, much below its martensitic transition temperature and study the growth of the equilibrium phase (martensitic). Here we have used a MSM alloy of composition Ni$_{45}$Co$_{5}$Mn$_{38}$Sn$_{12}$ and study the growth of equilibrium phase (martensitic) with time at 25K and 50K starting with different phase fractions of non-equilibrium austenite phase created by traversing different H-T paths. We show that, in standard process of cooling and measuring in 4 Tesla field the rate of growth of martensitic phase is drastically less at 25K compared to the 50K in spite of having almost similar starting fraction of the non-equilibrium austenite phase, indicating the effect of kinetic arrest of the first order transformation\cite{chatto, vks}. On the contrary, the growth rate is found to be almost similar at 25K and 50K (but significantly higher compared to the previous case) when the sample was cooled in 8T to the respective temperatures and the measurement field is isothermally reduced to 4T. Further, it is found that though the growth rate depends on the initial non-equilibrium phase fraction, it has very weak dependence on temperature for this protocol. It is shown that while cooling in 8T, nucleation of the martensitic phase is completely inhibited and the system approaches 50K or 25K with fully arrested austenite phase. The nucleation of the martensitic phase starts only while reducing the field. Whereas, while cooling in 4T, the nucleation of the martensitic phase has started at much higher temperature ($\approx$150K) and the system reached 25K or 50K with substantial fraction of transformed martensitic phase. Attempt is made to explain the intriguing time evolution through the motion of interface between the austenite and martensitic phase.     

\maketitle\section{Experimental Details}
The ribbons sample of MSM alloy of composition Ni$_{45}$Co$_{5}$Mn$_{38}$Sn$_{12}$ (Sn-12) was prepared by melt spinning of the pre-melted alloy prepared from high purity elements under high purity argon atmosphere. The composition and the crystal structure was determined from energy dispersive x-ray spectroscopy (EDXS) and powder x-ray diffraction (XRD) techniques respectively.  The details of the sample preparation and characterization are given in Ref.\cite{alok2}. The magnetization measurement was carried out in commercial 14T VSM (PPMS) made by Quantum Design, USA. For magnetic measurements, cooling or heating was always done at the rate of 1.5K/min. and the field changing was done at the rate of 100 Oe/sec. The time decay measurements were done immediately after temperature stabilization or field change. The M-H measurements are performed while sweeping the field at the same rate. 

\maketitle\section{Results and discussions}
Magnetization (M) as a function of T for Ni$_{45}$Co$_{5}$Mn$_{38}$Sn$_{12}$ is shown in Fig. 1a. After cooling the sample in zero field from 350K, 4T field was switched on at 5K and M is measured while warming for the zero-field cooled (ZFC) branch. Then M is measured while cooling in the same field for field-cooled cooling (FCC) and while warming after field cooling (FCW). The FCC and FCW branches show a paramagnet to ferromagnetic (austenite) transition at high-T followed by a hysteretic austenite to martensite first order transition at lower-T. Though the thermal hysteresis in M(T) closes around 65K, the bifurcation between ZFC and FCC (or FCW) branches indicates that the first order transition is not completed down to the lowest T. This is a clear indication of the presence of kinetically arrested high-T austenite phase, at low-T\cite{rawat,vks,ito}. In a recent study on a Ni-Mn-In-Co system it is shown that there is metastability while cooling across the first order austenite to martensitic transformation whereas there is no metastability while heating\cite{kustov}. This indicates that there is supercooling but no superheating in this system. This asymmetry between superheating and  supercooling is commonly observed in the case of melting of solids where nucleation occurs on the surface\cite{soibel}.
\begin{figure}[htbp]
	\centering
		\includegraphics[width=8cm]{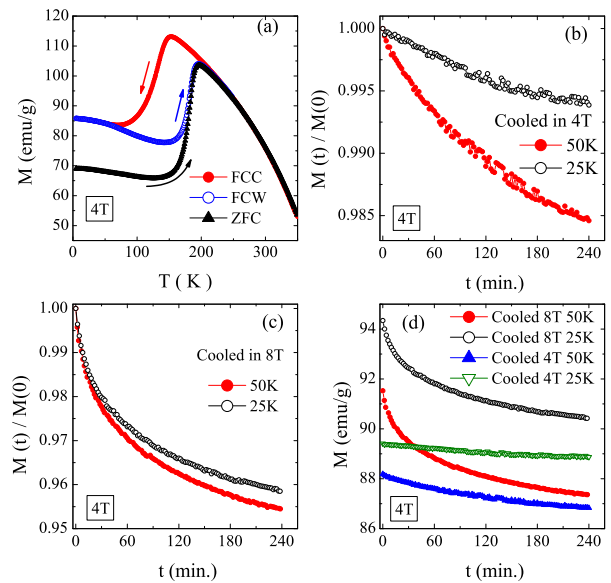}
		\caption{(Color online) Magnetization (M) as a function of temperature (T) and time (t) measured at 4 Tesla field following various protocols for Ni$_{45}$Co$_{5}$Mn$_{38}$Sn$_{12}$. (a) M vs. T for ZFC, FCC and FCW paths. (b) Normalized M vs. t at 25K and 50K after cooling from 350K in 4T. (c) Normalized M vs. t at 25K and 50K after cooling from 350K in 8T and isothermally reducing the field to measuring field of 4T at the respective temperatures. (d) Decay of absolute value of M as a function of t.}
\label{fig:fig1}
\end{figure}

To probe the low-T state, we measured M as a function of time (t), M(t), at 25K and 50K by reaching the respective measurement temperatures by two paths in H-T space. One by cooling from 350K in 4T down to 25K or 50K and measured M(t). Fig.1b shows the M(t) vs. t after normalization with the respective M(at t=0), M(0), for the respective measurement temperatures. In the second protocol, the same points in H-T space are reached by cooling from 350K in 8T down to 25K or 50K and then isothermally reducing to the measurement field of 4T. Fig. 1c shows the normalized M(t) vs. t for 25K and 50K. It may be noted that though both the measurement temperatures are below the closure of thermal hysteresis related to first order austenite to martensitic transformation, yet, the magnetic states at 25K or 50K are far from equilibrium. The decrease in M with t indicates that the low-M martensitic phase grows from the high-M metastable austenite phase fraction. 
However, this growth of martensitic phase has following rather intriguing features:

(1)  The rate of growth is higher for 50K compared to 25K for both the protocols. In the first protocol of measurement i.e. cooling and measuring in 4T, the decay in M at 50K is almost 3 times than that of 25K. This is contrary to the transformation process related to first order transition, in which case the barrier in free energy decreases as the system approaches the supercooling limit leading to faster growth of low-T equilibrium phase with the decrease in temperature. On the contrary, the observed trend in the growth rate is similar to the kinetically arrested state indicating that the austenite phase fraction is in non-equilibrium glass like state at these temperatures \cite{chatto, rawat, chaddah2}.

(2)  The decay in M is not directly related to the initial magnetization value as shown in Fig. 1d. Though the 4T cooled 25K measured state has intermediate starting value of M it shows the slowest decay.

(3)  The most intriguing aspect is the significantly different rates of growth of the equilibrium martensitic phase at the same temperature and measuring fields i.e. 4T and 25K or 50K. While in the 1st protocol (Fig. 1b), the magnetization decreases by 0.6$\%$ over 4 hours at 25K the same is about 1.6$\%$ for 50K. In contrast to that the magnetization decreases by more than 4$\%$ over 4 hours for both 25K and 50K in the 2nd protocol (Fig. 1c).

It was theoretically predicted and also experimentally verified for the vortex matter that the region of metastability depends on traversed paths in two variable space\cite{chaddah, roy2}. Though this may offer some justification to the observed drastically different relaxation rates for the two paths (point $\#$3), the slower relaxation at lower temperature (point$\#$1) cannot be explained only on the basis of first order transformation process. Since, for a first order transition, the free energy barrier height decreases with the decrease in temperature, as the system approaches the supercooling limit, the relaxation rate increases with decrease in temperature. However, observed behavior is a typical signature of the glass like arrested state. For a glassy system the relaxation rate becomes critically slow with the decrease in temperature as is also observed for the ``magnetic glass''\cite{chatto,chaddah2, vks}. 
\begin{figure}[htbp]
	\centering
		\includegraphics[width=8cm]{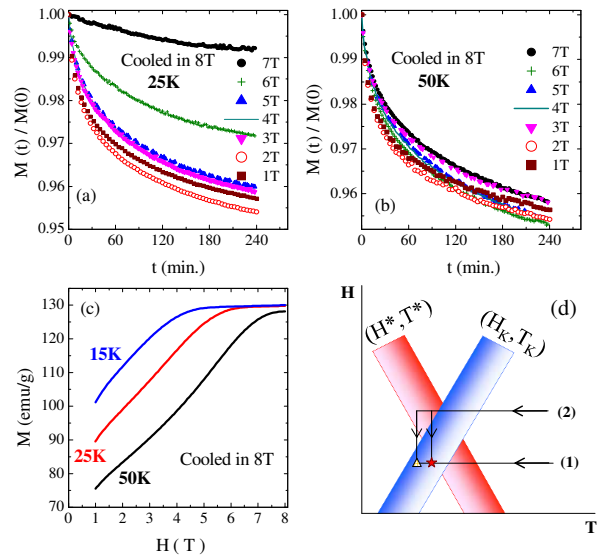}
		\caption{(Color online) Time and H dependence of M after cooling in 8T from 350K and schematic H-T diagram for Ni$_{45}$Co$_{5}$Mn$_{38}$Sn$_{12}$. (a) Normalized M vs. t at 25K after cooling from 350K in 8T and isothermally reducing to different measuring H. (b) M vs. t at 50K following the same protocol as (a). (c) M vs. H at 15K, 25K and 50K while reducing the field after cooling from 350K to respective measurement T in 8T. (d) Schematic H-T diagram depicting supercooling (H*, T*) and Kinetic arrest (H$_K$, T$_K$) bands and also the two prominent paths followed in the present study.}
\label{fig:fig2}
\end{figure}

To comprehend these observations we measured M(t) at 25K and 50K in different H after cooling from 350K in 8T and isothermally reducing the field to the measurement field at the respective measurement temperatures.  M(t) for 25K and 50K are shown in Fig. 2a and 2b respectively. While the growth rate of martensitic phase for all the measurement H $\leq$ 7T are high (more than 4$\%$ decay of M over 4hrs) at 50K (Fig. 2b), similar growth rate is observed only for the measuring H $\leq$ 5T at 25K (Fig. 2a). The growth rate of martensitic phase at 25K is rather low for 7T, less than 1$\%$ decay in M is observed over 4hrs, the same is higher but less than 3$\%$ for 6T measuring H. This apparently puzzling behavior appears to be related to the H at which the nucleation of the martensitic phase begins while reducing the field from 8T at the respective temperatures. Fig. 2c shows the isothermal field reduction M-H curves at 15K, 25K and 50K after cooling the sample in 8T. The sharp fall in M while reducing the H from 8T marks the onset of the nucleation process of the martensitic phase. At 50K the nucleation start around 7.5T and for 25K it is around 6T. This can be justified from the schematic H-T phase diagram shown in Fig 2d, which is similar to the phenomenological phase diagram proposed initially for high-T ferromagnetic phase (See Fig. 1a of Ref.\cite{kranti}). While cooling in 8T following path-2, a large fraction of supercooled austenite phase gets arrested because various regions encounters the kinetic arrest (T$_{K}$, H$_{K}$) band before crossing the corresponding supercooling limits (T*, H*). When the field is reduced at 50K or 25K, the system traverses the (T$_{K}$, H$_{K}$) band from the opposite side and the arrested phase fractions gets progressively de-arrested resulting in the nucleation of the martensitic phase. Since at 50K, the (T$_{K}$, H$_{K}$) band is encountered at a higher field than the 25K path, nucleation of the martensitic phase starts at a higher field for 50K. This explains the observed time evolution of M indicating the growth of martensitic phase at 25K and 50K for different fields (Fig. 2 a and b). The decrease in M is faster when the measuring field is lower than the field for onset of nucleation, above that field there is hardly any nucleation or time evolution of M and around this field there is intermediate rate of increase in the nucleation of the martensitic phase.

We attempt to unify the observed H-T path dependent diversity by quantitative analysis of the time dependent evolution of the equilibrium phase (martensite) from the non-equilibrium phase (austenite) at 25K and 50K for path-2. Let P$_{Eq}$(t) and P$_{Neq}$(t) be the equilibrium and non-equilibrium phase fractions at time `t' and saturation magnetization of the two states are denoted as M$_{Eq}$ and M$_{Neq}$ respectively.  If P$_{Eq}$(0) is the starting equilibrium phase fraction then P$_{Neq}$(0) = $\left[1-P_{Eq}(0)\right]$ is the starting non-equilibrium phase faction. 

Since broad distribution of relaxation rates leads to a logarithmic relaxation\cite{amir}, we consider a logarithmic time decay of the non-equilibrium austenite phase as 
\begin{equation}
	P_{Neq}(t) = P_{Neq}(0)\left[1- Dln(t/t_{0})\right]
\end{equation}
	where D is the rate constant.
It can be shown that the magnetization at any time t is given by
\begin{equation}
M(t) = M(0) - \left[1-P_{Eq}(0)\right]\Delta M Dln(t/t_{0}) 
\end{equation}
	where $\Delta$M = M$_{Neq}$ - M$_{Eq}$ and the starting magnetization value M(0), which can be shown to be
	 
M(0) = P$_{Eq}$(0) M$_{Eq}$ + $\left[1-P_{Eq}(0)\right]$ M$_{Neq}$ 

= M$_{Neq}$ - P$_{Eq}$(0)$\Delta$M.

	The zero field cooled state is fully converted martensitic phase whose M (vs. H) attains technical saturation around 1T and does not undergo reverse martensitic transformation even up to 14T field\cite{alok2}. Hence, the magnetization value at the respective fields (for H$>$1T) from the zero-field cooled M-H curves (not shown here) at the measurement temperatures (25K and 50K) can be taken as corresponding M$_{Eq}$. When the system is cooled in 8T it reaches 50K or 25K with fully arrested austenite phase, which is also shown to be soft ferromagnetic phase having technical saturation around 1T. Hence, extrapolation of the M-H curves of 25K or 50K (Fig. 2.c) to the respective measurement fields ($>$1T) gives the corresponding M$_{Neq}$ values. These values along with the corresponding starting measured magnetization values, M(0), are used to obtain the two phase fractions and thus $P_{Eq}$(0). The rate parameter D is the only fitting parameter for the fitting of  Eq. 2 to the M vs t data at 25K and 50K (Fig. 2 a and b). 
\begin{figure}[htbp]
	\centering
		\includegraphics[width=6.5cm]{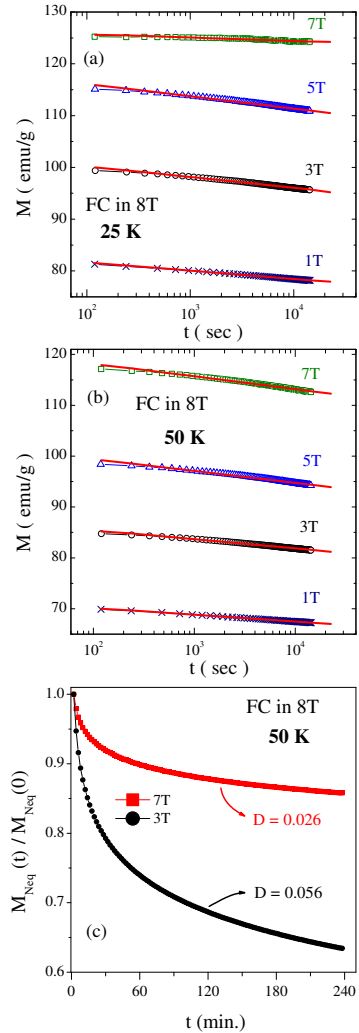}
		\caption{(Color online) Time dependence of M at 25K and 50K in different measuring H after cooling in 8T for Ni$_{45}$Co$_{5}$Mn$_{38}$Sn$_{12}$ . (a) and (b) Logarithmic time dependence of M as well as the one-parameter fitting to Eq. (2) at 25K and 50K respectively. Data and fitting for only the alternate H values are shown for clarity. (c) The calculated normalized non-equilibrium phase fraction vs. t at 50K for 3T and 7T after cooling in 8T. The data is taken from Fig. 2b and the rate parameter D is taken from the fitting to the Eq. 2.}
\label{fig:fig3}
\end{figure}
These fittings are shown in Fig. 3a and 3b (fittings for only the alternate field values are shown for clarity), which indicate that Eq. 2 gives a reasonable description (for t$>$100 Sec.) for the time evolution of the coexisting phases of this system. It may be noted that small deviation from the logarithmic behavior is experimentally observed in the relaxation of electrical resistance of a MSM alloy of Ni-Mn-In-Co system\cite{kustov}.  Moreover, it is shown from rigorous theoretical calculations that logarithmic relaxation is valid only in certain asymptotic limit and for this case it for the larger time (beyond 100 Sec.)\cite{amir, amir2}.  

The above mentioned quantitative analysis brings out some intriguing aspect as shown in Fig. 3c. Though the 3T and 7T curves show similar decay of total normalized magnetization at 50K (as shown in Fig. 2b), they have very different fractions of non-equilibrium phase. After cooling in 8T to 50K when the field is reduced to 7T, it has about 80$\%$ of non-equillibrium phase whereas at 3T non-equillibrium phase fraction is only about 25$\%$. Thus the normalized decay rate of the non-equillibrium phase fraction is significantly lower at 7T compared to 3T as shown in Fig. 3c. This indicates a significant dependence of the decay rate D on the starting non-equillibrium phase fraction. Hence, we plot the decay rate D as a function of the starting non-equilibrium phase fraction $\left[1-P_{Eq}(0)\right]$ in Fig. 4. It is rather significant that D appears to follows some kind of scaling with $\left[1-P_{Eq}(0)\right]$ which may have interesting consequences for broad classes of materials. However, decrease of D with increase in $\left[1-P_{Eq}(0)\right]$ as well as its very weak temperature dependence is a matter of real concern.

The non-equilibrium phase fraction is related to the ``degree of metastability" and for a first order transition this dictates the kinetics of the growth of the equilibrium phase\cite{landau}. In the initial stage of transition, the total volume of the nuclei of the equilibrium phase is rather small and their formation as well as growth is not correlated, since it is arising from the fluctuation in energy, hence, the effect on the ``degree of metastability" is rather limited. With larger volume fraction of the equilibrium phase, the nature of the growth process, that is now correlated, is very different. However, in the present case the nucleation and growth process of the equilibrium phase encounter two opposing effects as depicted by the overlapping supercooling and kinetic arrest bands (Fig. 2d). Notwithstanding this complication, it offers us significant control on process of nucleation by the second control parameter H and allows us to initiate nucleation at much lower temperatures by traversing different H-T paths. Recently, it is shown for a CMR manganite that even for same degree of metastability or same fraction of non-equilibrium phase the rate of growth depend on the H-T history \cite{arxiv}. This is attributed to the H-T path dependent critical radius of the nuclei leading to different sizes of mesoscopic domains whose distinct growth rate is dictated by the relaxation of the interfaces. The importance of the motion of the structural interface between two magnetic phases for the first order transition and its kinetic arrest leading to glass like state at low temperature is highlighted through an experimental study on another CMR manganite system\cite{sha}. Further, this study relates the motion of the interface of the structurally dissimilar phases with the growth of martensitic phase of shape memory alloys.
\begin{figure}[htbp]	\centering
		\includegraphics[width=8cm]{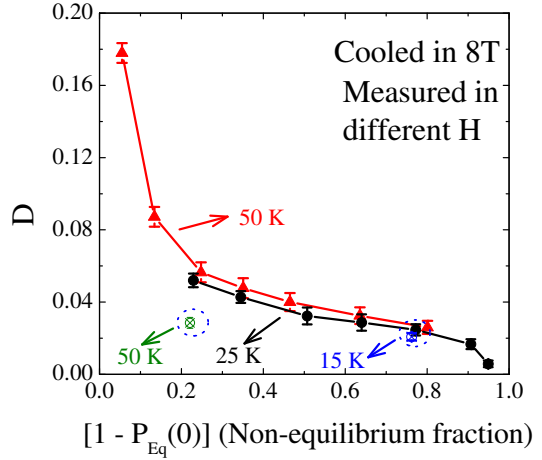}
		\caption{(Color online) Variation of decay rate (D) of Eq. (2) as a function of starting non-equilibrium fraction [1-P$_{Eq}$(0)] for different path in H-T space. The close circles (black) and up triangles (red) are derived from the fittings of Fig. 3a and 3b. The two points related to the fitting of M vs t by Eq. (2) for the following two others paths are also shown by open symbols with error bars encircled with dotted lines: (i) Cooling in 4T to 50K (green), (ii) Cooling in 8T to 15K, then isothermally reducing the field to 4T (blue).}
\label{fig:fig4}
\end{figure}  
  
In the present case, while cooling in 8T the nucleation is prevented by the process of kinetic arrest and it is initiated only at a much lower temperature by reducing the field. For example, while cooling in 4T, nucleation of martensitic phase starts $\approx$150K, whereas after 8T cooling it is started only at the measurement temperature of 50K or below. It is known that the critical radius R$_C$ for nuclei formation is dictated by the difference between the equilibrium transformation temperature ($\approx$200K) and the nucleation temperature\cite{landau,lif,chaik}. After cooling in 8T to the respective measurement temperature when field is reduced isothermally, then the de-arrest of the arrested austenite phase start below a certain H with the formation of the nuclei of the martensitic phase of critical radius R$_C$ which populate as the field is progressively decreased. Thus, the starting non-equilibrium phase fraction is inversely related to the number of martensitic nuclei when the system is subjected to the time relaxation keeping the field constant. We expect growth process is dictated by the interface area of the initial state\cite{sha}. This explains the observed scaling of the relaxation rate D with the non-equilibrium phase fraction of Fig. 4, for path-2. Observation of the almost same relaxation rate D for the same non-equilibrium phase fraction for 25K and 50K indicates that the R$_C$ is similar for both the temperatures and the D is governed only by the number of martensitic nuclei or the starting non-equilibrium phase fraction.  It is reasonable to expect that since the nucleation is prevented at the normal first order transformation process, critical radius has reached the lowest physical dimension and losses further temperature dependence. This conjecture is tested by measuring the time relaxation after cooling in 8T to 15K and isothermally reducing the field to 4T. It is rather significant that even at 15K the relaxation rate scales with the starting non-equilibrium phase fraction (calculated from 15K M-H of Fig. 2c) and merges with curves of 25K or 50K within the error bars as shown in Fig. 4. Thus, for path-2, it can be safely asserted that relaxation at or below 50K will follow the same scaling with the starting non-equilibrium phase fraction and fall on the same curve of Fig. 4. 
 
The above mentioned analysis is reconfirmed by calculating the relaxation rate for the 4T-cooled state at 50K. For this case, larger fraction of austenite phase has already converted to the equilibrium marteinsitic state while cooling in 4T and the nucleation has started $\approx$150K (as shown in Fig. 1a) having larger critical radius R$_C$. Thus this state has larger size domains of the martensitic phase compared to the state created by cooling in 8T and reducing the field at 50K. Hence, for this 4T-cooled state, the starting interface area is smaller for the same fraction of the non-equilibrium phase compared to the state crated by path-2, resulting in a significantly smaller decay rate as show in Fig. 4. The D value for this 4T-cooled state is much below the scaling curve of 50K or 25K (as shown in Fig. 4) and reinforces the above mentioned analysis. Moreover, this explains the reason behind the observed drastically different growth rates for path-1 and path-2 (Fig. 1). Hence, intense theoretical efforts backed by experiments on variety of such magnetic systems where arrested kinetics given rise to glass like long range ordered magnetic states are essential for deeper understanding of the apparently intriguing observations presented here. Further, such path dependent metastability is expected to be rather ubiquitous and needs to be investigated in other materials identified as ``magnetic glasses'' (see for example Ref.\cite{rav})

\maketitle\section{Conclusion}
We show that low-T metastable magnetic states in a shape memory alloy system, Ni$_{45}$Co$_{5}$Mn$_{38}$Sn$_{12}$, decay with very different rates at the same measuring field and temperature depending on how that H and T are reached from high temperature side. In this system the H-T induced broad first order magnetic transition is inhibited by arrest of transformation kinetics resulting in persistence of high-T austenite phase fraction which coexist at low-T with equilibrium martensite phase. The fractions of these phases depend on the H-T paths; however, magnetization of the metastable coexisting phase shows glass like decay at low-T. This study reveals that the decay rate of the metastable coexisting phase scales with the starting non-equilibrium phase fraction but is almost independent of temperature when the decay is studied after isothermal field reduction at a temperature much below the closure of the hysteresis related to the first order transition or much below the martensitic transformation process. Whereas, the decay rate is significantly small at the same H and T when the standard process of cooling and measuring in the same field is followed. This apparently anomalous behavior is explained on the basis that the nucleation starting much below the 1st order transition temperature have smallest possible critical radius R$_C$ of the nuclei of the martensitic phase and consequently larger interface area for the same fractions of phases compared to the case when the nucleation is started at a much higher temperature having much larger R$_C$. The long time relaxation at low temperature takes place with the growth of the respective nuclei (or domain of the martensitic phase) and is related to the available starting interface area. Further, this study indicates that when the process of nucleation is inhibited by the arrest of kinetics allowing nucleation at much lower temperature by reduction of H, then the size of the critical radius is independent of the temperature.  

\maketitle\section{Acknowledgment}
C.X. and R.V.R. acknowledge support from National Research Foundation, Singapore through the CREATE program on Nanomaterials for Energy and Water Management.

\end{document}